\documentclass[12pt]{article}
\usepackage{latexsym}

\def\lesssim{{\
\lower-1.2pt\vbox{\hbox{\rlap{$<$}\lower5pt\vbox{\hbox{$\sim$}}}}\ }} 
\def\gtrsim{{\
\lower-1.2pt\vbox{\hbox{\rlap{$>$}\lower5pt\vbox{\hbox{$\sim$}}}}\ }}

\begin{document}

\begin{titlepage}

\begin{flushright}
    September 1997
\end{flushright}

\vspace{0.5cm}

\begin{center}
{\Large\bf Domain Wall Problem of Axion and Isocurvature Fluctuations 
in Chaotic Inflation Models} 
    
\vspace{1.5cm} 
{\large S. Kasuya$^{a}$, M.~Kawasaki$^{a}$ and
T.~Yanagida$^{b}$}\\
\vspace{1cm}
{\it $^{a}$Institute for Cosmic Ray Research, University of Tokyo, 
Tokyo 188, Japan\\
$^{b}$Department of Physics, University of Tokyo, Tokyo 133, Japan}

\end{center}

\vspace{2.0cm}

\begin{abstract}
We study the domain wall problem of an axion in chaotic inflation 
models.  We show that the production of domain walls does not occurs 
if the Peccei-Quinn scalar has a flat potential and its breaking 
$F_{a}$ is larger than $\sim 10^{15}$GeV. We find that too large 
isocurvature fluctuations are produced for such high $F_{a}$.  In 
order for those isocurvature fluctuations to be consistent with 
observations of the large scale structure of the universe, the 
self-coupling constant $g$ and the breaking scale of the Peccei-Quinn 
scalar should be $g \simeq (1-2.8)\times 10^{-13}$ and $F_{a} \simeq 
(0.6-1.5)\times 10^{15}$GeV, respectively.  In particular the value of 
self-coupling constant is almost the same as that required for a 
chaotic inflation, which strongly suggests that the Peccei-Quinn 
scalar itself is an inflaton.
\end{abstract}


\end{titlepage}
\newpage


The axion~\cite{Peccei,Wilczek,Kim,DFS} is the most natural solution
to the strong CP problem in QCD~\cite{tHoot}. The axion is associated
with breaking of the Peccei-Quinn symmetry whose breaking scale $F_a$
is stringently constrained by laboratory experiments, astrophysics and
cosmology; the allowed range of $F_a$ is between $10^{10}$~GeV and
$10^{12}$~GeV~\cite{Kolb-Turner} in the standard cosmology. The lower
limit comes from consideration of the cooling of the supernova SN1987A
by axions and the upperbound is obtained by requiring the the cosmic
axion density should not exceed the critical density of the universe.
Thus the axion is one of attractive candidates for dark matter if
$F_a$ takes higher values in the allowed region.

However, in the standard cosmology the axion has a domain-wall
problem~\cite{Sikivie}. Since the potential of the axion which is
formed through QCD instanton effects has $N$ discrete minima ( $N$:
color anomaly factor), domain walls are produced at the QCD scale.
The domain walls with $N > 1$ form very complicated network together
with axionic strings which are produced associated with spontaneous
breaking of Peccei-Quinn symmetry and do not disappear in the
subsequent evolution of the universe, which leads to a cosmological
disaster.\footnote{
The domain walls with $N=1$ have disk-like structure whose boundaries
are axionic strings, and they collapse and disappear so rapidly that
they are cosmologically harmless.}

In the inflationary universe~\cite{Guth,Sato} it is generally expected 
that the domain wall problem is solved since the exponential expansion 
during inflation makes the phase of Peccei-Quinn scalar (=axion) 
homogeneous beyond the present horizon of the universe and hence 
axions settle down to the same minimum of the potential in the entire 
universe.  However, in the chaotic inflationary 
universe~\cite{Linde1} the domain walls are produced through large 
quantum fluctuations of axions generated in the chaotic 
inflation~\cite{Lyth1}.  The quantum fluctuations of axions $\delta a$ 
are given by $H/(2\pi)$ where $H$ is the Hubble constant during 
inflation.  $\delta a$ is $10^{12-13}$GeV for chaotic inflation and it 
is larger than the Peccei-Quinn scale $F_a$.  This means that 
the axion phases $\theta_a \equiv a/F_a$ become random after inflation 
and domain walls are produced in the same way as the standard 
cosmology.

Furthermore, the quantum fluctuations of axions cause anisotropies of 
the CMB~\cite{Turner,Lyth2,Linde2} which are too large.  Since the 
axion is massless during inflation, the axion fluctuations do not 
contribute to the fluctuations of the total energy density of the 
universe, i.e.  isocurvature fluctuations.  After the axion acquires a 
mass $m_a$ at the QCD scale, the axion fluctuations become density 
fluctuations given by $\delta \rho_a/\rho_a \sim \delta 
\theta_a/\theta_a$, which cause CMB temperature fluctuations $\delta 
T/T \sim \delta \theta_a/\theta_a$.  Since $\delta \theta_a \sim 
O(1)$, the CMB anisotropies produced are $O(1)$, which contradicts 
observations~\cite{COBE}.

It was pointed out in ref.\cite{Linde2,LiLy} that the above two 
problems are simultaneously solved if the potential of the 
Peccei-Quinn scalar is very flat.  For a flat potential the 
Peccei-Quinn scalar $\Phi_a$ can have a large expectation value $\sim 
M_{\rm pl}$ at the epoch of inflation.  Then we should take $\langle 
\Phi_a \rangle$ as the effective Peccei-Quinn scale ``$F_{a}$'' 
instead of $F_a$, and the phase fluctuations are suppressed.  
Therefore, the production of domain walls is suppressed and 
isocurvature fluctuations decrease.  However, in the previous 
work~\cite{KKY} we showed that the Peccei-Quinn scalar field 
oscillates after inflation and violently decays into axions through 
parametric resonance~\cite{KLS}, which results in large phase 
fluctuations of $O(1)$ and hence formation of domain walls.

Recently, the detailed numerical calculations on the formation of 
topological defects through parametric resonance~\cite{KK} were made 
and it was found that the critical breaking scale below which the 
topological defects are formed is $\sim 10^{15}$GeV. Thus, the domain 
walls are produced for the standard Peccei-Quinn scale ($\sim 
10^{12}$GeV), and only the axion models with $N=1$ are allowed as is 
pointed in ref.~\cite{KKY}.  However, if we take the Peccei-Quinn 
scale larger than $\sim 10^{15}$GeV, we can avoid the domain wall 
problem for general axion models with $N>1$.  Therefore, in this 
letter, we investigate the axion domain wall problem for $F_a \gtrsim 
10^{15}$GeV. We show that the domain wall and isocurvature problems 
are indeed solved for axions with $F_a \sim 10^{15}$GeV. It is also 
shown that such axions produce significant amount of isocurvature 
fluctuations and anisotropies of the cosmic microwave background 
radiation (CMBR) which can be detected by future satellite 
experiments.

First let us consider the evolution of the Peccei-Quinn scalar
$\Phi_a$ which has a flat potential given by
\begin{equation}
      V(\Phi_{a}) = \frac{g}{4}(|\Phi_{a}|^2 - F_{a}^2)^2
      = \frac{g}{4}(\phi_{a}^2 - F_{a}^2)^2,
      \label{potential}
\end{equation}
where $\phi_{a}$ is the radial part of $\Phi_{a}$, defined as 
\begin{eqnarray}
    \Phi_{a}(x) & \equiv &  \phi_{a}(x)
    \exp(i a(x)/|\langle\phi_a\rangle|),\\
    & & \phi_a(x): {\rm real},~ -\infty < \phi_a < \infty, \nonumber \\
    & & a(x):{\rm real},~ -
    \pi/2 \le a(x)/|\langle\phi_a\rangle| \le \pi/2. \nonumber 
\end{eqnarray}
We also assume an inflaton field $\chi$ with a potential:
\begin{equation}
    V(\chi) = \frac{\lambda}{4} \chi^4,
    \label{inflaton-pot}
\end{equation}
where the coupling constant $\lambda$ is about $10^{-13}$ which is 
required to produce the anisotropies of CMBR observed by 
COBE~\cite{COBE}.  In the chaotic condition of the early universe both 
$\phi_{a}$ and $\chi$ have large expectation values greater than the 
Planck scale. During inflationary epoch both fields slowly roll down 
with relation given by
\begin{equation}
    \phi_{a} \simeq \left(\frac{\lambda}{g}\right)^{1/2}\chi.
    \label{phi-chi-relation}
\end{equation}
The inflationary epoch ends when $\chi \simeq M_{p}/3$ ($M_{p}$: 
Planck mass) and $\chi$ and $\phi_{a}$ begin to oscillate.  The 
potential energy of the Peccei-Quinn scalar $\phi_{a}$ at this epoch 
is much larger than the height of the potential hill ($=gF_{a}^4/4$).  
Thus $\phi_{a}$ oscillates beyond the potential hill.  Since the 
potential of $\phi_{a}$ has minima at $\phi_{a} = \pm F_{a}$, 
$\phi_{a}$ settles down to $F_{a}$ or $-F_{a}$.  If the initial 
fluctuations of $\phi_{a}$ are large, the final values of $\phi_{a}$ 
are different at different places in the universe, which results in 
the formation of domain walls.  In the present case, the fluctuations 
of $\phi_{a}$ comes from quantum fluctuations during inflation and 
their amplitude is given by 
\begin{equation}
    \delta \phi_{a} \simeq \frac{H}{2\pi} \simeq 
    \frac{\sqrt{\lambda}\chi^2}{\sqrt{6\pi}M_{p}} \simeq
    \frac{8\sqrt{2}\sqrt{\lambda}M_{p}}{\sqrt{3\pi}},
    \label{delta-phi}
 \end{equation}
where we take $\chi\simeq 4M_{p}$ because cosmological scales (1kpc -- 
3000Mpc) correspond to the event horizon lengths when $\chi \sim 
4M_{p}$. Thus, at the end of inflation $t_i$ ($\phi_{a}(t_i)\simeq 
\sqrt{\lambda/g} M_{p}/3$), the fluctuations of $\phi_a$ amount to 
\begin{equation}
    \frac{\delta \phi_{a}}{\phi_{a}}(t_i)\simeq 11 \sqrt{g}.
    \label{delta-phi/phi}
\end{equation} 
At the critical epoch $t_{*}$ when the Peccei-Quinn scalar settles
down to one of two minima of the potential (i.e. $\phi_a \sim F_a$),
the change $\Delta A$ of the amplitude $A$ of $\phi_a$ per one
oscillation is given by
\begin{equation}
    \frac{\Delta A}{A} \sim \frac{H_{*}}{\omega}(t_*)
    \sim \frac{\sqrt{g}F_{a}}{\sqrt{\lambda}M_{p}},
    \label{delta-A/A}
\end{equation}
where $\omega$ is the frequency ($\sim \sqrt{g}F_{a}$), $H_{*}$ is the
Hubble constant at the critical epoch. Here we have assumed that the total
energy density is always dominated by the inflaton and radiations
produced through the decay of the inflaton.  If the dynamics is pure
classical and the initial fluctuations $\delta \phi_{a}/\phi_{a}$ is
less than $\Delta A/A$, it is expected that $\phi_{a}$ settles down to
the same minimum and no domain walls are formed. ( This criterion for 
domain wall production is confirmed by numerical calculation in
ref.~\cite{KK}.)  From eqs.  (\ref{delta-phi/phi}) and
(\ref{delta-A/A}), the condition for no domain wall production is written as
\begin{equation}
    F_{a} \gtrsim 15\sqrt{\lambda}M_{p} 
    \simeq 4\times 10^{13}{\rm GeV}.
    \label{Fa-const}
\end{equation}
Notice that this condition is independent of the coupling $g$.
However, the actual dynamics is not pure classical because the
oscillating $\phi_{a}$-field decays into $\phi_{a}$-particles and
axions through parametric resonance~\cite{KKY}. Thus the condition
(\ref{Fa-const}) is modified if we take account of the $\phi_a$ decay.
In fact, the effect of the parametric resonant decay on the formation
of topological defects has been investigated in ref.~\cite{KK}, and it is 
shown that the above condition
(\ref{Fa-const}) becomes more stringent as
\begin{equation}
    F_{a} \gtrsim 6\times 10^{14}{\rm GeV}.
    \label{Fa-const2}
\end{equation}

Since the above Peccei-Quinn scale is much higher than 
the usual values (i.e. $\sim 10^{12}$GeV), we should care about cosmic 
density of the axion. The density parameter $\Omega_{a}$ of the axion 
is related by $F_{a}$ as~\cite{Kolb-Turner}
\begin{equation}
    \Omega_{a}h^2 \simeq 7.9\times 10^{2}
    \left(\frac{F_{a}}{10^{15}{\rm GeV}}\right)^{1.18}\theta_{a}^2,
    \label{omega}
\end{equation}
where $h$ is the present Hubble constant in units of 100km/sec/Mpc.
Since the fluctuations of $\theta_{a}$ are small in the present model,
$\theta_{a}$ is almost homogeneous over the entire universe. Thus we can take
$\theta_{a}$ in eq.(\ref{omega}) as a free parameter. Assuming that
the axion is dark matter ($\Omega_{a} h^2 \simeq 0.25$), we obtain
\begin{equation}
    \theta_{a} \simeq 0.018
    \left(\frac{F_{a}}{10^{15}{\rm GeV}}\right)^{-0.59}.
    \label{theta}
\end{equation}

Next, we consider the isocurvature fluctuations of the axion.  The 
fluctuations of the axion $\delta a$ and the effective Peccei-Quinn 
scale ``$F_{a}$'' during inflation are given by $H/(2\pi)$ and 
$``F_{a}"=\phi_{a}$, respectively.  Since the axion phase $\theta_{a}$ 
is related to $a/\phi_{a}$, the isocurvature fluctuations of the axion 
density ($= F_{a}^2 \delta \theta_{a}^2/2$) are given by 
\begin{equation}
    \delta_{\rm iso} \equiv 
    \left(\frac{\delta \rho_{a}}{\rho_{a}}\right)_{\rm iso}
    = \frac{H}{\pi \phi_{a} \theta_{a}}
    \simeq \sqrt{\frac{2}{3\pi}}\frac{g^{1/2}\chi}{M_{p}\theta_{a}},
    \label{iso}
\end{equation}
where we use eq.(\ref{phi-chi-relation}). 

On the other hand, the inflaton generates adiabatic fluctuations
which amount to
\begin{equation}
    \delta_{\rm ad} \equiv 
    \left(\frac{\delta \rho_{a}}{\rho_{a}}\right)_{\rm ad}
    = \frac{2H^3}{3\pi V'} \simeq 
    \frac{4\sqrt{2\pi}}{9\sqrt{3}} \frac{\lambda^{1/2}\chi^3}{M_{p}^3}, 
    \label{ad}
\end{equation}
at horizon crossing (i.e. when wavelengths become equal to the horizon
length).  Then the ratio $\alpha$ of $\delta_{\rm iso}^2$ to
$\delta_{\rm ad}^2$ is written by~\cite{KSY}
\begin{equation}
    \alpha \equiv \frac{\delta_{\rm iso}^2}{\delta_{\rm ad}^2}
    = \frac{81}{16\pi^2}
    \frac{gM_{p}}{\lambda \chi^4 \theta_{a}^2} 
    \simeq 2.0 \times 10^{-3}\theta_{a}^{-2} \frac{g}{\lambda}. 
    \label{ratio}
\end{equation}
The definition of $\alpha$ is the same as given in ref.~\cite{KSY} and
$\alpha$ is the ratio of the power spectra  of isocurvature and
adiabatic fluctuations. Using eq.(\ref{theta}), $\alpha$ is given by
\begin{equation}
    \alpha \simeq 6.4 (g/\lambda)
    \left(\frac{F_{a}}{10^{15}{\rm GeV}}\right)^{1.18}
    \gtrsim 3.5 (g/\lambda),
    \label{ratio2}
\end{equation}
where the inequality comes from eq.(\ref{Fa-const2}).  Notice that $g 
\ge \lambda \simeq 10^{-13}$ if the Peccei-Quinn scalar is not an 
inflaton.  On the other hand, if $\Phi_a$ is an inflaton, $g = \lambda 
\simeq 10^{-13}$.  Thus $g \ge 10^{-13}$ for any case.  The upperbound 
on $g$ is obtained from observations of large scale structure of the 
universe.  The ratio $\alpha$ should be less than about 10, otherwise 
the isocurvature fluctuations become so large that the predicted power 
spectrum contradicts observations of large scale structure~\cite{KSY}.  
This gives constraints on $g$ and $F_a$.  The allowed range of $g$ is 
$(1-2.8)\times 10^{-13}$, and $F_a$ should take $ (0.6- 1.5)\times 
10^{15}$GeV which is close to scales of grand unifications.  
Furthermore, since $g$ is almost the same as $\lambda$, it is natural 
to consider the Peccei-Quinn scalar an inflaton itself (i.e.  $g 
=\lambda$).

The mixture of isocurvature and adiabatic fluctuations is 
astrophysically interesting
because it gives a better fit to  the observations of the
large scale structure of the universe~\cite{KSY} than pure adiabatic
fluctuations in standard cold dark matter (CDM) scenario. In the
standard CDM scenario, the density fluctuations at scales of galaxies
and clusters are too large if the power spectrum $P(k)$ is normalized
to the COBE data. For example, the amplitude of mass fluctuations at
$8h^{-1}$Mpc, $\sigma_8$ is 1.4 for standard CDM with $h=0.5$, while
observed values of $\sigma_8$ are 0.57 from galaxy cluster
surveys~\cite{White-Efstathiou-Frenk}, $0.75$ from galaxy and cluster
correlations~\cite{Peacock-Dodds} and $0.5-1.3$ from peculiar velocity
fields~\cite{Strauss-Willick}. However, since isocurvature
fluctuations give six times larger contribution to CMBR anisotropies
at COBE scales, the mixture of isocurvature fluctuations decreases the
amplitude of the matter fluctuations and $\sigma_8$ is reduced to
$\simeq 1 - 0.5$ for $\alpha = 1-10$~\cite{KSY}.\footnote{
On the other hand, if fluctuations is pure isocurvature, $\sigma_8$ is 
about 0.1 which contradicts observations.}

Furthermore, anisotropies of CMBR in the present model can be
distinguished from those produced by pure adiabatic fluctuations
because the shape of power spectrum of CMBR anisotropies is quite
different from that of pure adiabatic fluctuations at small angular
scales.  Since direct searches for the axion are impossible for $F_{a}
\gg 10^{12}$GeV, the observations of the CMBR anisotropies by future
satellite experiments are very crucial to test the present model.

In summary, we have studied the domain wall problem of axion in 
chaotic inflation models and showed that the production of domain 
walls is suppressed if the Peccei-Quinn scalar has a flat potential 
and its breaking $F_{a}$ is larger than $\sim 10^{15}$GeV. We have 
also found that the observable amount of the isocurvature fluctuations 
is produced for such high $F_{a}$.  The present model is consistent 
with observations if the self-coupling constant and the breaking scale 
of the Peccei-Quinn scalar are $g \simeq (1-2.8)\times 10^{-13}$ and 
$F_{a} \simeq (0.6-1.5)\times 10^{15}$GeV. In particular the value of 
self-coupling constant is almost the same as that required for an 
inflaton, which suggests that the Peccei-Quinn scalar itself is the 
inflaton.


\begin{thebibliography}{99}
\bibitem{Peccei} R.D. Peccei and H.R. Quinn, 
    Phys. Rev. Lett. {\bf 38}, 1440 (1977).
\bibitem{Wilczek} 
  F. Wilczek, 
  Phys. Rev. Lett. {\bf 40}, 279 (1978).
\bibitem{Kim} 
  J.E. Kim, 
  Phys. Rev. Lett. {\bf 43}, 103 (1979);\\
  M. Shifman, A. Vainshtein and V. Zakharov, 
  Nucl. Phys. {\bf B166}, 493 (1980).
\bibitem{DFS} 
  M. Dine, W. Fischler and M. Srednicki, 
  Phys. Lett. {\bf  B104}, 199 (1981).
\bibitem{tHoot} G. 't Hooft, 
  Phys. Rev. Lett. {\bf 37}, 8 (1976).
\bibitem{Kolb-Turner}
  E.W. Kolb and M.S. Turner,
  The Early Universe, Addison-Wesley, (1990).
\bibitem{Sikivie} P. Sikivie,
  Phys. Rev. Lett. {\bf 48}, 1152 (1982).
\bibitem{Guth} A.H. Guth,
  Phys. Rev. {\bf D23}, 347 (1981).
\bibitem{Sato} K. Sato,
  Mon. Not. R. astr. Soc. {\bf 195}, 467 (1981).
\bibitem{Linde1} A.D. Linde,
  Phys. Lett. {\bf B129}, 177 (1983).
\bibitem{Lyth1}
  D.H. Lyth,
  Phys. Rev. {\bf D45}, 3394 (1992);\\
  D.H. Lyth and E.D. Stewart,
  Phys. Rev. {\bf D46}, 532 (1992).
\bibitem{Turner} 
  M.S. Turner and F. Wilczek,
  Phys. Rev. Lett. {\bf 66}, 5 (1991);\\
  M. Axenides, R. Brandenberger and M.S. Turner,
  Phys. Lett. {\bf B126}, 178 (1983);\\
  P.J. Steinhardt and M.S. Turner,
  Phys. Lett. {\bf  B129}, 51 (1983);\\
  D. Seckel and M.S. Turner,
  Phys. Rev. {\bf D32}, 3178 (1985);\\
  L.A. Kofman and A.D. Linde,
  Nucl. Phys. {\bf B282}, 55 (1987).
\bibitem{COBE} G.F. Smoot {\it et al.},
  Astrophys. J. {\bf 396}, L1 (1992).
\bibitem{Lyth2} D.H. Lyth,
  Phys. Lett. {\bf 236}, 408 (1990).
\bibitem{Linde2} A.D. Linde,
  Phys. Lett. {\bf B259}, 38 (1991).
\bibitem{LiLy} A.D. Linde and D.H. Lyth,
    Phys. Lett. {\bf B246}, 353 (1990).
\bibitem{KKY} S. Kasuya, M. Kawasaki, and T. Yanagida,
  hep-ph/9608405.
\bibitem{KLS} L. Kofman, A.D. Linde and A. Strarobinsky,
    Phys. Rev. Lett. {\bf 73}, 3195 (1994);\\
    L. Kofman, A.D. Linde and A. Strarobinsky,
    Phys. Rev. Lett. {\bf 76}, 1011 (1996);\\
    D. Boyanovsky, H.J.de Vega, R. Holman, D.-S. Lee and A. Singh,
    Phys. Rev. \textbf{D51}, 4419 (1995);\\
    M. Yoshimura,
    Prog. Theor. Phys. \textbf{94}, 873 (1995);\\
    Y. Shtanov, J. Traschen and R. Brandenberger,
    Phys. Rev. \textbf{D51},5438 (1995);\\
    S.Yu. Khlebnikov and I.I. Tkachev,
    Phys. Rev. Lett. \textbf{77}, 219 (1996);\\
    S. Kasuya and M. Kawasaki,
    Phys. Lett. {\bf B 388}, 686 (1996). 
\bibitem{KK} S. Kasuya and M. Kawasaki,
  hep-ph/9703354.
\bibitem{KSY} M. Kawasaki, N. Sugiyama and T. Yanagida,
  Phys. Rev. {\bf D 54}, 2442 (1996). 
\bibitem{White-Efstathiou-Frenk} S.D.M. White, G. Efstathiou and 
  C. Frenk,
  Mon. Not. R. astr. Soc. {\bf 262}, 1023 (1993).
\bibitem{Peacock-Dodds} J.A. Peacock and S.J. Dodds,
  Mon. Not. R. astr. Soc. {\bf 267}, 1020 (1994).
\bibitem{Strauss-Willick} M.A. Strauss and J.A. Willick, 
  Physics Reports {\bf 261}, 271 (1995).
\end{thebibliography}
\end{document}